\documentclass{article}%
\usepackage{amssymb}
\usepackage{amsfonts}
\usepackage{amsmath}%
\usepackage{graphicx}
\providecommand{\U}[1]{\protect\rule{.1in}{.1in}}
\begin{document}

\title{Reality and the Probability Wave}
\author{Daniel Shanahan\\PO Box 301, Raby Bay, Queensland, Australia, 4163}
\maketitle

\begin{abstract}
Effects associated in quantum mechanics with a divisible probability wave are
explained as physically real consequences of the equal but opposite reaction
of the apparatus as a particle is measured. \ Taking as illustration a
Mach-Zehnder interferometer operating by refraction, it is shown that this
reaction must comprise a fluctuation in the reradiation field of complementary
effect to the changes occurring in the photon as it is projected into one or
other path. \ The evolution of this fluctuation through the experiment will
explain the alternative states of the particle discerned in self interference,
while the maintenance of equilibrium in the face of such fluctuations becomes
the source of the Born probabilities. \ In this scheme, the probability wave
is a mathematical artifact, epistemic rather than ontic, and akin in this
respect to the simplifying constructions of geometrical optics.\smallskip
\smallskip

\textbf{Keywords \ }probability wave $\cdot$ measurement problem $\cdot$ self
interference $\cdot$ Schr\"{o}dinger's cat $\cdot$ Mach-Zehnder interferometer
$\cdot$ local realism

\end{abstract}

\section{Introduction}

It seems to have been Max Born who first referred to the waves of probability
(\textit{Wahrscheinlichkeitswellen}) which released from the usual binds of
reality have provided an elegant and for practical purposes highly successful
explanation of the phenomenon of self interference (Born \cite{born}).

Yet these probability waves are no less mysterious than the mystery they were
invoked to explain. \ It can hardly be doubted that as a particle passes
through an interferometer, such as the Mach-Zehnder of Fig. 1, something
wave-like must be moving through both arms. \ But\ I will contend here for an
ontologically less interesting explanation of this wave-like effect. \ 

I will assume that the particle itself follows a single and well-defined path
at all times, and that its choice of that path is determined, not in the
intrinsically probabilistic manner supposed by standard quantum mechanics
(SQM), but in accordance with underlying physical microprocesses. \ Insisting
on the strict and local application of laws of conservation, I will show that
any change in the wave characteristics of the particle as it adopts that path
must be accompanied by a wave-like disturbance of equal but opposite effect in
the scattering medium of the apparatus.\ 

As each particle is measured, there will be thus two wave systems evolving
through the experiment, precisely coordinated but of complementary effect
-\ the wave-like particle, and the correspondingly wave-like response of the
apparatus to the scattering of that particle. \ My contention will be that as
this response propagates through the experiment, it mimics the presence of a
further version or versions of the particle itself\medskip\medskip

\ \ \ \includegraphics[width=8.0cm]{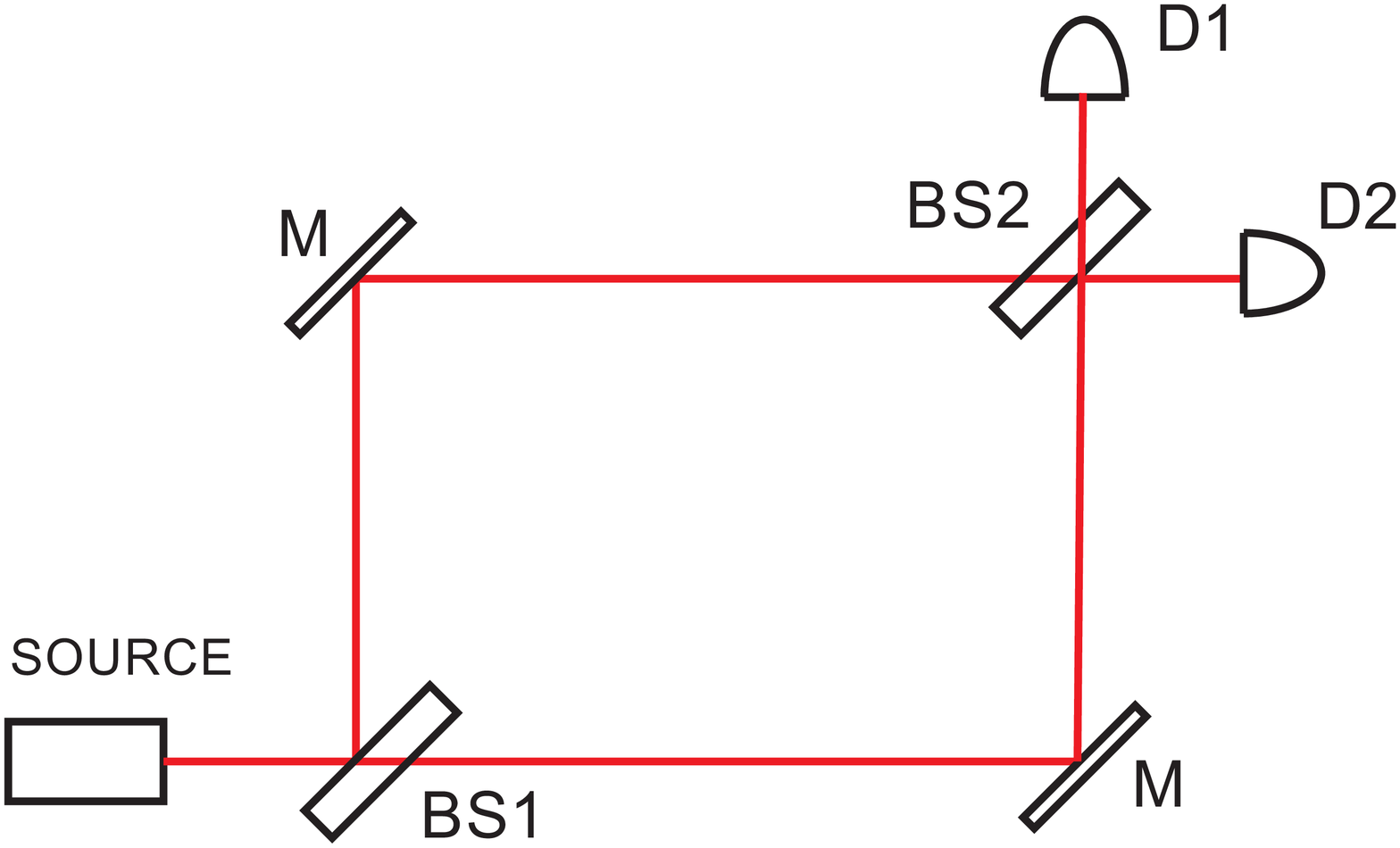}\medskip\medskip

\begin{quotation}
\textbf{Fig. 1} \ A Mach-Zehnder interferometer: \ In standard quantum
mechanics, the probability wave divides at beam splitter $BS1$ and self
interference occurs as the partial waves recombine at $BS2$. \ As now
proposed, the scattering of a photon into one or other arm at $BS1$ induces by
reaction a fluctuation in the reradiation field, and it is this that
interferes with the photon at $BS2$.\medskip\medskip
\end{quotation}

This response will constitute a microscopic imbalance in the scattering medium
of the apparatus. \ The tendency of the apparatus to resist the accumulation
of imbalance will be the source of the Born probabilities. \ The evolution of
the imbalance through the experiment will explain the seemingly probabilistic
alternative states that have been discerned in quantum measurement.

The difficulties that have arisen in SQM from the notion of an
observer-instigated wave function collapse are well illustrated by
Schr\"{o}dinger's \textit{reductio} of the unobserved cat that is at once dead
and alive (Schr\"{o}dinger \cite{schrodinger}), and by the iteration of the
problem of the cat in the paradox of Wigner's friend (Wigner \cite{wigner}).
\ Those difficulties will be avoided here, not merely by obviating the need
for an observer, as is the aim of, for example, the objective collapse and
many worlds reinterpretations of quantum mechanics, but by eliminating the
Schr\"{o}dinger evolution of alternative probabilistic states that has been
thought to necessitate that collapse.

In developing the argument, I will refer to simple beam splitters operating by
refraction and to a Mach-Zehnder interferometer assembled from such beam
splitters. \ Despite the mysteries of\ the quantum, the phenomenon of
refraction is itself relatively non-controversial, and by concentrating on the
refraction of photons, I may avoid the suggestion of \textquotedblleft new
physics\textquotedblright, and here at least, the complications of the de
Broglie wave, of which something has been said elsewhere (Shanahan
\cite{shanahan1} and \cite{shanahan2}). \ 

The Mach-Zehnder is not the famous instance of self interference. \ But this
conceptually simple interferometer will allow the demonstration of an illusion
- the apparent ability of an indivisible particle to be in two places at once
- \ an illusion induced, as I will show, by conservation, quantization, and
the wave-like nature of the elementary particles.

\section{Conservation and measurement}

At the level of the quantum, a measurement apparatus does not measure any
individual particle. \ It must proceed by forcing (projecting) the particle
into one or other of the eigenmodes defined by the apparatus, typically
eigenmodes of some property to be measured. \ In the limit of large numbers, a
beam of particles tends to separate between those modes so as to conserve that
property in the measured beam, this being, as will be discussed in Sect. 4,
the basis of the Born probabilities. \ By processing a sufficient number of
particles in this way, and guided by those probabilities, something may thus
be learned of the components of the property in the original beam.

But unless the incident particle was already in an eigenmode of the apparatus,
conservation is not observed in the particle itself. \ Indeed it was at one
time proposed on high authority (see Bohr, Kramers\ and Slater \cite{bks})
that conservation must be merely approximate or \textquotedblleft
statistical\textquotedblright\ in microscopic processes. \ That suggestion was
withdrawn following experimental confirmation of the conservation of momentum
in scattering processes (the Bothe \cite{bothe}\ and Compton \cite{compton}
experiments). \ Conservation is now more usually regarded as a meta-principle
against which an otherwise promising proposal might be judged and found
wanting. \ Certainly, a close attention to conservation has proved crucial on
occasion to the understanding\ of quantum phenomena (see, for instance,
Bloembergen \cite{bloembergen}).

Yet in according roles in measurement to chance and nonlocality, SQM seems
careless of the conserved properties of physics. \ The energy of a
superposition of waves, and thus of interacting particles, is determined by
their relative phase and degree of overlap. \ It is not explained how energy
is to be conserved if such a superposition evolves discontinuously or
nonlocally. \ \ The arbitrariness of such an evolution is difficult to
reconcile with the symmetries contemplated by Noether's theorems, and would
seem to deny the local conservation and continuity supposed by the gauge
principles of modern field theories (see, for instance, Ryder \cite{ryder},
Chap. 3)

I will assume in this paper, not only that laws of conservation are observed
exactly in quantum measurement, but that movements in the properties conserved
develop through the process of measurement in the local and deterministic
manner that was supposed by classical physics. \ From that assumption, I will
show that if a strict accounting is kept of those movements, the notion of a
collapsing probability wave becomes redundant to the operation of conserved
microprocesses. \ 

Consider the interaction of a single particle with an ideal 50:50
beamsplitter. \ According to SQM, the probability wave associated with the
particle (in more formal terms, the wave function or state vector) divides at
the beam splitter in accordance with conservation as,%
\begin{equation}
\left\vert \psi_{in}\right\rangle ->\frac{1}{\sqrt{2}}(\,\left\vert \psi
_{I}\right\rangle +\left\vert \psi_{II}\right\rangle \,),\label{probdiv}%
\end{equation}
where $\left\vert \psi_{in}\right\rangle $ is the amplitude of the probability
wave of the incoming particle, while $\left\vert \psi_{I}\right\rangle $ and
$\left\vert \psi_{II}\right\rangle $ are those of the partial waves exiting
ports $I$ and $II$ respectively.\ 

In dividing in this way, the probability wave has entered the so-called
Schr\"{o}dinger phase in which each measurement possibility (component of the
wave function) evolves through the experiment in a local and deterministic
manner until the occurrence of an observation (a measurement). \ According to
the collapse postulate of SQM, the wave function then collapses in an
intrinsically probabilistic, discontinuous and nonlocal manner into one or
other of the two possible measurement outcomes ($\left\vert \psi
_{I}\right\rangle $ or $\left\vert \psi_{II}\right\rangle $).\ 

If conservation is to continue following this collapse, there must be a
reaction of equal but opposite effect to the change in the particle, and if
this reaction is to accord with local causality, it must occur as and where
the change in the particle is effected, that is to say, in the
apparatus\footnote{It might be thought that the role of the apparatus in
measurement was settled by von Neumann in his acclaimed \textit{Mathematische
Grundlagen der Quantenmechanik} \cite{neumann} \ \ But the "pointer" states of
the apparatus contemplated there by von Neumann were correlated with the
changed state of the particle rather than, as in the response of the apparatus
considered here, anticorrelated with the change in the particle. The
significance of the pointer in von Neumann's analysis (see in particular
his\ Chap. VI) was that it is some such visual indication, rather than the
microscopic particle that, by impinging on the consciousness of the observer,
collapses the wave function (according to von Neumann).}. \ On including this
reaction, Eqn. (\ref{probdiv} ) becomes,%

\begin{equation}
\psi_{in}->\frac{1}{\sqrt{2}}[(\left\vert \psi_{I}\right\rangle +\Delta
A_{I})+(\left\vert \psi_{II}\right\rangle +\Delta A_{II})],\label{comb}%
\end{equation}
where $\Delta A_{I}$ is the change that would occur in the apparatus if the
photon were to exit port $I$, and $\Delta A_{II}$ is the corresponding change
if it were to exit port $II$.

This response by the apparatus must compensate for what is lost by the
particle in one component of the measured property and gained in the other.
\ In a 50:50 beamsplitter, and if the particle takes path $I$, the response
may be expressed formally as,
\begin{equation}
\Delta A_{I}=\frac{1}{\sqrt{2}}(\left\vert \psi_{II}\right\rangle -\left\vert
\psi_{I}\right\rangle ),\label{respI}%
\end{equation}
\ and if it takes path $II$,%
\begin{equation}
\Delta A_{II}=\frac{1}{\sqrt{2}}(\left\vert \psi_{I}\right\rangle -\left\vert
\psi_{II}\right\rangle ),\label{respII}%
\end{equation}
where in either case the negative sign denotes a reduction in amplitude or
equivalently a phase opposed to that of the particle\footnote{From the
symmetry under time reversal of Maxwell's equations, it may be argued,
following Stokes (see Hecht \cite{hecht}, chap. 4.6.3 and 4.10), that the
component of the response of the medium propagating in the same channel as the
particle must be of opposite phase to that particle.}. \ 

With the reaction of the apparatus now brought into the account, self
interference can be explained without recourse to the convoluted rigmarole of
division and collapse supposed by SQM. \ It is only necessary to assume that
as the particle encounters the scattering medium of the beam splitter, it does
not enter the probabilistic superposition supposed by SQM, but is forced
immediately toward one or other of the two exit ports of the apparatus. \ 

As this occurs the reaction of the apparatus, evolving through the apparatus
in the same manner as the photon has two effects. \ Along the path not taken
by the photon, it mimics the effect of a divided photon. \ Along the path that
is taken by the photon, it diminishes the effect of the photon so as to
complete the illusion of a divided photon. On the particle exiting port $I$,
the result must be, from Eqn. (\ref{respI}),%

\[
\left\vert \psi_{I}\right\rangle +\Delta A_{I}=\frac{1}{\sqrt{2}}(\left\vert
\psi_{I}\right\rangle +\left\vert \psi_{II}\right\rangle ),
\]
and on its exiting path $II$, likewise from Eqn. (\ref{respII}), \
\[
\left\vert \psi_{II}\right\rangle +\Delta A_{II}=\frac{1}{\sqrt{2}}(\left\vert
\psi_{I}\right\rangle +\left\vert \psi_{II}\right\rangle ),
\]
thus avoiding all necessity for the division and collapse supposed by SQM. \ 

And that is what will be supposed here - that particle and apparatus adopt
their measured states in a local and causal manner as the one interacts with
the other, rather than belatedly and retrospectively following the collapse of
the wave function, a collapse that could in principle occur much later, or in
the absence of particle detectors, never at all.

The argument has relied crucially, of course, on the assumption that the
response of the apparatus evolves along the same paths and in the same
wave-like manner as the particle itself. \ That this must be so is suggested
by the detailed and continuing requirements of conservation. \ SQM assumes
that the probability wave divides in accordance with conservation. \ If
conservation is to continue following collapse, when only one of the two
possible measurement outcomes has been realized, the response of the apparatus
must supply what is missing and must continue to do so as the system evolves.

However, nothing has yet been said of the microprocesses underlying this
wave-like response. \ I will consider in the next section a class of
scattering processes, ubiquitous in quantum measurement, in which there can be
no doubt that the response of the medium does indeed propagate though the
experiment in the same wave-like manner as the particle itself.

\section{The response of the medium}

Before confronting the mysteries of measurement directly, it will be
instructive to consider the response of a scattering medium where the
scattered particle is allowed no choice of path - where there is no suggestion
of wave function collapse and accordingly no mystery at all.

One such case is refraction within an isotropic dielectric (such as glass) in
a\ region remote from discontinuity (see, for instance Born and Wolf
\cite{bornwolf}, Chap. II, and for an intuitive treatment, Feynman, Leighton
and Sands \cite{feynman}, Vol. I, Ch. 31). \ 

The interaction is entirely between the field of the photon and the charged
particles of the medium. \ If there were no charges within the medium, the
photon would pass entirely unaffected. \ In a dielectric, these are bound
charges, and the process is thus mediated by moments, primarily electric
dipole moments, induced by the flux of photons in the molecules of the
material. \ As each photon passes through the medium, it interacts with a vast
number of these molecules, driving in each molecule the oscillating divergence
of positive and negative charge distributions that is the source of its dipole
moment. \ 

In the semi-classical modelling of refraction - in which the incident wave is
continuous and only the medium is quantized - each induced moment is
approximated as an harmonic oscillator, essentially an oscillating electron
constituting a small electric dipole,%
\begin{equation}
\mathbf{p}=-\frac{q^{2}}{m\omega^{2}}\mathbf{E},\label{dipole}%
\end{equation}
where $\mathbf{E}$ is the incident electric field, $\omega$ is its frequency,
and $q$ and $m$ are the charge and mass of the electron respectively (see Born
and Wolf\textbf{\ }\cite{bornwolf}, chap. 2.4.1).

Keeping in mind an array of such dipoles, we need to consider a beam that
arrives, not continuously, but episodically in a flux of discrete photons.
\ Consistently with what is known of its interactions, the photon can be
regarded as a microscopic and, in the present context, essentially
indivisible, electromagnetic field having the form of a transverse wave.

As each dipole seeks to regain its unexcited state, it reradiates and does so,
potentially at least, in all directions. \ If a photon were to interact with a
single isolated dipole, it could be scattered in any of a range of possible
directions. \ But because the array of moments acquires its spatial
distribution of phase and amplitude from the inducing flux, this reradiation
(also referred to as the polarization field) interferes constructively only in
the direction of propagation of the flux itself (see, for instance, Barron
\cite{barron}, pp. 124-5). \ It is the composition of this induced
polarization field with the field of the photon that causes the change in
wavelength, and thus phase velocity, that is the origin of the refractive index.

Essentially this same process - the interference between a particle and a
secondary field that the scattering of that particle has itself induced - will
be identified in Sect. 6 as the explanation of the self interference observed
in a Mach-Zehnder interferometer. \ According to this view, refraction is
itself an elementary form of self interference, but one in which incident and
induced fields are\ able to interfere constructively in only one direction. \ 

The process of refraction is considerably more complicated than might be
supposed from the brief sketching above. \ Even if consideration could be
limited to the passage of just one photon, the field felt by each moment will
include reradiation from every other moment with which the photon is
interacting, and these moments will likely have one of more resonance
frequencies, and include contributions from higher order and magnetic moments.
\ In practice refraction and reflection are dealt with, not in terms of
microscopic properties, but the more easily measured macroscopic properties,
including in particular the refractive index $n$ and the bulk electric,
magnetic and polarization fields $E$, $B$ and $P$, and more will be said of
this in Sect. 5.

But I will ignore all such complications here (see instead the cited texts).
\ The essential points were made above and are in summary:

(a) \ that the interaction of photon with medium is solely with the charges of
the medium, which has the important consequence that the reaction of the
medium to any change in the photon is also mediated solely by those charges; and,

(b) that accompanying each photon, there\ is indeed another electromagnetic
field - the reradiation or polarization field - which having acquired its wave
characteristics and direction of propagation from the inducing flux is well
adapted to interfere with the photon or an accompanying photon.

Without as yet allowing a photon a choice of path, let us now introduce the
complication that the medium is birefringent. \ We will suppose that the
photon is propagating horizontally through a uniaxial crystal having its optic
axis aligned so that horizontal and vertical ($H$ and $V$) components of the
electric field of the photon induce moments of differing strengths, and thus
experience correspondingly different changes of wavelength and phase velocity.

The photon is not disrupted by these competing effects. \ It is able to
accommodate the differing phase velocities of its $H$ and $V$ components by a
continuing variation in its state of polarization as it passes through the
medium\footnote{A photon linearly polarized at $\theta$ to the vertical
evolves through the various stages of elliptical polarization until the
optical paths of its $H$ and $V$ components have differed by $2\pi$ at which
point it will have regained its original state of polarization, and the
sequence recommences. \ If, when the photon exits the medium, the paths of $H$
and $V$ components differ by $2n\pi+\pi$, the medium will have acted as a half
wave plate, if $2n\pi+\pi/2$, as a quarter wave plate.}. \ And here again,
there is the interference between photon and induced field that was
categorized above as self interference. \ Yet there has been no reason to
suppose that the process is in any way discontinuous, nonlocal or
probabilistic. \ 

But there is now an additional effect of some consequence. \ With the change
in polarization, a torsional reaction occurs in the medium. \ It is known from
the experiment of Beth in 1935 (Beth \cite{beth}) and the exploitation of the
Beth effect in optical traps and the like (see, for instance, Ashkin et al
\cite{ashkin}) that photons experiencing a change in polarization when
refracted by a dielectric target, not only impart linear and angular momentum
to that target, but may do so to the extent of causing observable movement of
the target.

Now consider refraction as it occurs in measurement, as when a photon
encounters the birefringence of a polarizing beamsplitter or the partially
reflective surface of a simple non-polarizing beamsplitter. \ The interaction
is again mediated solely by induced moments, but there are now alternative
paths of constructive interference available to the photon, and competing
influences on its characteristic structure that cannot be accommodated by a
mere change of wavelength or polarization. \ If the photon were freely
divisible it would separate between those paths in the manner supposed of the
continuous wave of classical physics (and of the probability wave of SQM).
\ But (at these energies and in this medium) the photon is indivisible,
and\ must adopt in its entirety one or other of the two available paths.

We have come at last to the crux of the argument. \ As the photon is forced
into one or other path, conservation (or equivalently Newton's third law)
demands that there be an equal but opposite reaction in the apparatus, a
reaction mediated solely by those moments with which the photon is interacting
- primarily moments in a narrow skein of molecules at and near a surface of
discontinuity within the scattering medium. \ 

The reaction of the apparatus can only comprise a fluctuation in the relative
strengths of the components of those moments and in the reradiation from those
components.\ \ And I stress again that, having acquired its distribution of
phase and amplitude from the inducing photon, this fluctuation in the
reradiation field will be constrained to propagate along the same paths of
constructive interference as those available to the photon itself. \ 

Ignoring losses, and assuming the ideal 50:50 beamsplitter of the previous
section, conservation demands that this fluctuation in the polarization field
have the form described formally by Eqn. (\ref{respI}) or Eqn. (\ref{respII}).
\ It will comprise a reduction in the field in the mode taken by the photon
(or a fluctuation of opposite phase to the photon) and an increase in that
field in the direction not taken. \ 

As contemplated in Sect. 2, there will thus be physically real waves
propagating in each path from the beam splitter.

\section{The Born probabilities}

How might a photon choose between these paths, and why should the beam divide
in apparent compliance with the Born rule?\ \ In a simple form appropriate to
discrete measurement outcomes, the rule says that,%
\begin{equation}
prob(a_{i})=\left\vert u_{i}\left\vert \psi\right\rangle \right\vert
^{2}\label{born}%
\end{equation}
where,
\[
\sum\left\vert u_{i}\left\vert \psi\right\rangle \right\vert ^{2}=1,
\]
and $prob(a_{i})$ is the probability that a particle in the state $\psi$ will
be found with the eigenvalue $a_{i}$, for which the corresponding eigenstate
is $u_{i}$ (Born \cite{born1}).

The assumption in SQM that the $prob(a_{i})$ are intrinsic to the particle
measured is taken to mean that the outcome of measurement is governed by pure
chance, or as it is said, \textquotedblleft irreducible quantum
randomness\textquotedblright\ (see, for instance, Khrennikov \cite{khrennikov}%
, Ch. 7). \ This species of probability seems to have been encouraged,
initially at least, by the apparently random nature of atomic transitions, but
it is only in its consistency with the idea of a divisible probability wave
that it seems indispensable to SQM. \ If it were to be admitted that the
particle was in\ a particular channel of the apparatus prior to its
observation, there could be nothing in the other channel to explain the self
interference observed at measurement - or so at least it has been assumed in
SQM. \ 

However, it is a significant clue to the real nature of these probabilities
that they lead, in the limit of large numbers, to conservation of the measured
property, or when there is no measured property, to some other balancing
consistent with conservation. \ That is so at least in those cases involving
alternative paths from a scattering event that lead to the self interference
of interest here. \ 

For example, the division effected by a polarizing beam splitter is consistent
with conservation of the components of polarization of the incident beam,
while that by a non-polarizing beam splitter operating by partial reflection
maintains the balancing of electromagnetic fields defined by the Fresnel
relations (of which something more will be said in the next section). \ 

The division of the beam will also be constrained by quantization.\ \ For
instance, in the measurement of spin $1/2$ by a Stern-Gerlach magnet, it is
only the component of spin in the direction of the field that is measured, but
quantization requires that particles adopt an alignment that is either spin up
or spin down with respect to that field. \ Conservation of angular momentum
then requires that a beam with its spin at\ $\varphi$ to the field divide
between spin up and spin down modes in the proportions, $\cos^{2}\varphi/2$
\ to $\sin^{2}\varphi/2$, in accordance with the trigonometric relation,
\[
\cos\varphi=\cos^{2}\frac{\varphi}{2}-\sin^{2}\frac{\varphi}{2}.
\]
For spin greater than $1/2$ the number of modes available is greater, but the
probabilities and the manner in which they transform on rotation are always
consistent with the conservation of angular momentum.

It is also possible to see in these examples why the Born rule expresses
probabilities as the squares of amplitudes. \ The conserved property is the
component of a wave or oscillation and thus a vector, which explains the
relevance to the Born probabilities of Pythagoras theorem and Hilbert space
and the continuity in the transformation of waveforms relied upon by Gleason's
theorem (Gleason \cite{gleason}).

It might thus be thought that conservation is \textit{the} explanation of the
Born probabilities. \ But conservation alone cannot explain these
probabilities. \ In whatever manner, a beam might divide, laws of conservation
will be satisfied by the equal but opposite reaction of the apparatus, and
this would be so even if the beam were to adopt in its entirety a single mode
of the measured property. \ But unless the incident beam was already in that
single mode, such a result is not observed, and it is only necessary to
consider why that is so to come to what is, I suggest, the true explanation of
the probabilities.{}

If the beam were to divide in any manner inconsistent with conservation, this
would involve a sustained transfer of measured property from beam to
apparatus, and a consequent reduction of entropy, contrary to the second law
of thermodynamics. \ In the case of a polarization beam splitter, for example,
such a transfer would induce a torsional strain in the apparatus, capable in
principle of doing work.

This suggests that the degree of conservation observed in the measured beam is
merely incidental to a process of recovery or equilibrium in the apparatus.
\ In such a process, each photon would \textquotedblleft
choose\textquotedblright\ its path, not by chance, but as determined by its
own particular circumstances, including the state of imbalance in which the
photon finds the scattering medium. \ 

It is not the case, as seems to be implied by the notion of intrinsic
probability, that the apparatus encountered by one particle is in the same
state as that met by the next. \ Unobserved macroscopically, various changes
of significance are occurring in the apparatus. \ The medium is in thermal
equilibrium and superimposed on that equilibrium is the fluctuating imbalance
induced by preceding measurements. \ Created in the path of following and
accompanying photons, this imbalance is eminently adapted to influence by
interference the ensuing beam. \ 

The Beth effect, referred to in Sect. 3, showed that such an imbalance\ is not
simply passed on without local effect to the wider environment, but may be
sufficiently enduring to cause observable dynamic change in the scattering
medium. \ But unlike a beam splitter, the suspended waveplate of Beth was what
might be termed a single-mode device, allowing no possibility of maintaining
equilibrium by a division of the incoming beam. \ 

The division by pure chance assumed by SQM would be an inferior, indeed
unreliable, way of maintaining that equilibrium. \ The measurement of any one
particle would as likely increase as decrease a pre-existing imbalance in the
medium. \ The variance of a distribution based on chance (that of the random
walk) increases with run-time, and in the tails of such a distribution an
excursion from balance could be substantial (see Mandel and Wolf
\cite{mandel}, Chap. 2.10.2). \ Indeed, any such departure would question the
compliance of the system with the second law. \ While it may be convenient to
model excursions from equilibrium in terms of the random walk, they tend in
practice to be self-limiting whatever the run-time.

Such a rebalancing may be less effective when particles arrive, not in a
steady beam, but singly, one after the other. \ Presumably there must come a
point at which a beam is so attenuated that the imbalance induced by one
particle will have dissipated before the arrival of the next. \ But although
experiments with attenuated beams have been reported, and some degree of
attenuation is necessary for the observation of self interference, it does not
appear to have been demonstrated that the Born probabilities survive in a beam
that is so attenuated that each measurement is in effect a separate
experiment. \ 

Finally here, something should be said of those situations in which SQM
supposes a probabilistic superposition of states, but there has been no
preliminary scattering into alternative paths, and accordingly no apparent
opportunity for the rebalancing contemplated above. \ In these situations, SQM
has simply assumed that if a particle, or even it might seem, a macroscopic
object such as the cat of Schr\"{o}dinger, could be in one state or another
state, and it is not known which, it must be in a probabilistic superposition
of such states. \ 

Schr\"{o}dinger asked us to imagine\ a cat confined within a box with a small
sample of a radioactive substance so positioned that if an atom of the
substance were to decay, it would cause a flask of hydrocyanic acid to shatter
and kill the cat (Schr\"{o}dinger \cite{schrodinger}). \ Until the box is
opened and its contents observed, it would have to be assumed (according to
SQM) that each atom is in a superposition of decayed and undecayed states,
implying in turn a superposition of the dead cat with the living cat. \ 

For Schr\"{o}dinger, this superposition of cats was illustrative, as he said,
of the \textquotedblleft quite ridiculous cases\textquotedblright%
\ (\textit{ganz burleske F\"{a}lle}) that might arise from the
observer-dependent collapse assumed by SQM. \ Yet notwithstanding its
apparently random nature, there is nothing in a radioactive decay to suggest
the self interference that encouraged the notion of an intrinsically
probabilistic superposition. \ There is no suggestion that the atom is in two
places at the same time or that its later state is interfering with its
earlier state. \ There is accordingly no compelling reason, other than
consistency with those more troublesome cases that do lead to self
interference, to suppose that the timing of such a decay is governed by pure
chance, rather than deterministic microprocesses as yet unknown. \ 

The mischief here is that this notion of intrinsic probability is likely to be
inhibiting investigation into the possibility that some deterministic
mechanism is indeed ordering the timing of such decays. \ 

\section{A beam splitter}

I consider how this process of rebalancing might work in a simple
non-polarizing beam splitter operating by partial reflection. \ (Two such beam
splitters will be needed for the Mach-Zehnder interferometer to be considered
in the next section). \ 

While it may not be feasible to follow the details of what is happening at the
quantum level, the fluctuation in fields that results from the passage of an
individual photon can be treated as a microscopic perturbation of the
continuous and macroscopic wave supposed by classical physics.

The relative amplitudes of reflected and refracted beams were deduced by
Fresnel from an elastic wave theory in 1823, (see Silverman \cite{silverman},
pp. 228-231), and given what is essentially their modern textbook derivation
by Lorentz in 1875 by insisting that Maxwell's equations be satisfied across
the inter-medial boundary (Lorentz \cite{lorentz}). \ The macroscopic wave is
assumed to divide so that the forces on the charges of the medium, whether
arising from incident or induced fields, are in a state of balance (as
contemplated in the preceding section). \ 

The continuity of Maxwell's equations across the boundary requires (assuming a
wave passing from medium $1$ to medium $2$ through a boundary in the
$xy$-plane) that,
\begin{align}
\left(  \varepsilon_{0}\mathbf{E}_{1}\mathbf{+P}_{1}\right)  _{z} &  =\left(
\varepsilon_{0}\mathbf{E}_{2}\mathbf{+P}_{2}\right)  _{z},\nonumber\\
\left(  \mathbf{E}_{1}\right)  _{xy} &  =\left(  \mathbf{E}_{2}\right)
_{xy},\label{cond}\\
\ \ \ \mathbf{B}_{1}\  &  \mathbf{=}\ \mathbf{B}_{2}.\nonumber
\end{align}
where $\mathbf{E}$, $\mathbf{B}$ and $\mathbf{P}$ are respectively the
macroscopic electric, magnetic and polarization fields and $\varepsilon_{0}$
is the permittivity of free space (see, for example, Feynman \cite{feynman},
Vol. II, chap. 33).

In considering the quantized wave, the Fresnel relations can be taken to
define, not the steady state contemplated by Lorentz, but that notional point
of equilibrium about which the system fluctuates as photons are variously
reflected or refracted from the semi-reflective boundary. \ The fields must
remain in balance and for this to occur there must be a continuing
readjustment, not of the boundary conditions themselves, but of the manner in
which those conditions are satisfied. \ 

Consider, for example, the first of conditions (\ref{cond}), which is obtained
by asserting, in the $z$-direction, Coulomb's law, which in dielectric form
is,%
\[
\nabla\cdot\mathbf{E}=-\frac{\nabla\cdot\mathbf{P}}{\varepsilon_{0}}.
\]

On the side of the boundary to which a photon departs, there will be\ (as
compared with the steady state supposed classically) a fleeting increase in
the electromagnetic field supplied by the beam, and on the other side of the
boundary, a corresponding decrease in that field. \ This fluctuation will
induce by reaction compensating changes in the dispositions and relative
strengths of the components of moments and in the corresponding components of
the ambient polarization field. \ 

Whether the photon is\ reflected or transmitted, the fields at the boundary
will thus remain continuous, but at the cost of a microscopic departure from
the state of balance defined by the mean intensity of the incident beam.
\ This fluctuation in fields will influence in turn the choice of path made by
a following particle, thus contradicting the assumption of SQM that
measurement is intrinsically probabilistic. \ 

SQM effectively suppresses these fluctuations in field strengths by invoking
the probability wave, a phenomenon having more in common with the
constructions of geometrical optics than the quantized wave with which SQM
must ultimately deal. \ Having invoked this probability wave, SQM has had to
contrive its collapse, and thus the need for a measurement and an observer
that led to the \textquotedblleft measurement problem\textquotedblright. \ 

It is not until that collapse occurs that SQM gives effect to the quantization
of the measured beam, but by then the continuity of fields assumed by
Maxwell's equations has been lost in discontinuity and nonlocality. \ 

\section{The Mach-Zehnder interferometer}

Consider again the Mach-Zehnder interferometer of Fig. 1. The interference at
$BS2$ is now between real waves, these being\ the photon and the secondary
wave that was generated by reaction at $BS1$ as the photon was forced to adopt
one or other path through the interferometer. \ As discussed in Sect. 5, this
fluctuation in the polarization field maintains microscopically the continuity
and balancing of fields supposed classically by Maxwell's equations and the
Fresnel relations. \ 

As in SQM, each set of waves recombining at $BS2$ will have originated from
the scattering of a single photon at $BS1$. \ For photons of sufficiently like
frequency, the phase difference $\Delta$ between the two paths will thus
remain substantially the same from one photon to the next. \ Thus\ no matter
how attenuated or incoherent the original beam from the source may have been,
the recombining waves will demonstrate observable interference at the second
beam splitter. \ 

Let us suppose that $BS1$ and $BS2$ are non-polarizing lossless $50:50$
beamsplitters so constructed and aligned that when the upper and lower optical
paths to detector $D1$ differ by $\Delta$, the corresponding paths to detector
$D2$ will differ by $\Delta+\pi$. \ If $\Delta=0$, the recombining waves will
interfere constructively toward $D1$, but destructively in the direction of
$D2$. \ The photon will favour the path that preserves the integrity of its
waveform. \ Photons scattered at $BS2$ will thus register only at $D1$. \ If
$\Delta=\pi$, those photons will register instead at $D2$. \ In either case
the result will coincide with the prediction of SQM.

If the recombining waves are neither entirely in nor out of phase, the beam
will divide at $BS2$ in accordance with the intensities determined by the
interference occurring in $BS2$\footnote{For detection at $D1$, for instance,
and assuming recombining waves of phases $\chi_{1}$ and $\chi_{2}$, the
intensity (probability) is.%
\begin{align*}
\left\vert e^{i\chi_{1}}+e^{i\chi_{2}}\right\vert ^{2}  & =\left[  \left(
e^{i\left(  \chi_{1}+\chi_{2}\right)  /2}\right)  \left(  e^{i\Delta
/2}+e^{-i\Delta/2}\right)  \right]  ^{2},\\
& \thickapprox\cos^{2}\frac{\Delta}{2},
\end{align*}
where $\Delta=\chi_{1}-\chi_{2}$, and the first factor on the right hand side
in the first line\ has been equated with unity.}. \ In SQM, these intensities
correspond to the Born probabilities and are,%
\begin{equation}
\frac{prob(D1)}{prob(D2)}=\frac{\cos^{2}\frac{\Delta}{2}}{\sin^{2}\frac
{\Delta}{2}},\label{prob}%
\end{equation}
where $\Delta$ is again the difference between the two optical paths.

To provide a physically realistic explanation of these probabilities, I
briefly repeat what will now be an all too familiar refrain. \ The photon is
indivisible and must adopt in its entirety one or other path. \ As it does,
whatever force or effect is ensuring that indivisibility, must induce by
reaction in $BS2$, an imbalance in the dipole moments mediating the
interaction. \ If the photon takes the path toward $D1$, the coherent merger
of photon and secondary wave in that direction induces in $BS2$, a reaction of
energy,
\begin{equation}
\sin^{2}\frac{\Delta}{2},\label{proba}%
\end{equation}
tending to bias the ensuing flux toward $D2$. \ Conversely, coherence in the
direction of $D2$ must induce an imbalance,
\begin{equation}
\cos^{2}\frac{\Delta}{2},\label{probb}%
\end{equation}
creating a bias toward $D1$. \ From (\ref{proba}) and (\ref{probb}), the
maintenance of equilibrium within $BS2$ will ensure an (approximate) division
between $D1$ and $D2$ in the proportions predicted by the Born probabilities
((Eqn. \ref{prob}) ).

There is no suggestion in the above that the secondary wave is itself in any
sense a photon or part of a photon. \ Essentially, it is a fluctuation in the
polarization field capable of survival over the time frame of the experiment
and having an equal but opposite effect to the change occurring in the photon itself.

\section{Conclusion}

As described in this paper, it is an illusion - literally in the Mach-Zehnder,
a trick done with mirrors - that the photon is in two places at once. \ As the
indivisibility of the photon constrains it to one path,\ the response of the
apparatus, evolving through the experiment in the same wave-like manner as
the\ photon, creates the impression that the photon is somehow occupying both
paths. \ And with waveforms in each path, it is only natural that if an
attempt is made to locate the particle within those paths,\ the visibility of
the interference is diminished accordingly.

What then is the probability wave? \ Were it not now so commonplace, it might
seem a wondrous thing that an image can be propagated with such fidelity from
a reflective surface or through a pane of glass. \ The feat is no less
impressive when it is realized that the path of each photon is determined by
the fluctuating states of a vast multitude of mutually interfering
electromagnetic fields, these being the field of the photon itself, those of
accompanying photons, reradiation from moments induced by all those photons,
and further reradiation induced by the original reradiation.

In this multitude of interacting waveforms, there is no single wave, nor even
a divided wave, that is identifiable with the simply constructed probability
wave of SQM. \ Even if measurement were in some degree intrinsically
probabilistic, the probability wave could be no more than a mathematical
convenience, epistemic rather than ontic, and similar in this respect to the
constructions of geometrical optics. \ Such a construction may suggest where
the particle is likely to go, but not why it must goes there.

There are other interpretions of quantum mechanics in which the particle is
assumed to have, as it does here, a definite position and velocity, and hence
a well-defined trajectory. \ These include de Broglie's pilot wave theory of
1928 \cite{pilot}, and the rediscovery and revision of that theory by Bohm in
1952 (Bohm \cite{bohm1} and \cite{bohm2}, and see the review by Goldstein
\cite{goldstein1}). \ Also in this category is the recent proposal by Gao, who
argues for the physical reality of the particle from the implications of
protective measurement (see, for instance, Aharonov et al \cite{aharonov}),
and shows how the wave function, and in turn the Born probabilities, might be
defined by the rapid and discontinuous motion of the particle in question (Gao
\cite{gaowave}).

The theory presented here departs crucially from those proposals, as it does
from other interpretations of quantum mechanics, in identifying the mysterious
wave-like effect in the path (or paths) not taken by the particle as the
consequence of well-understood and otherwise familiar underlying
microprocesses. \ To explain self interference, the present proposal thus has
no need for a novel ontology, such as those interesting but inevitably
contentious ontologies that have characterized other theories. \ 

I have concentrated here on the Mach-Zehnder and not discussed at all the
better known double-slit experiment. \ But with its paths diverging to
macroscopic separations, the Mach-Zehnder provides I suggest the more
compelling illustration of self interference.

Nor of course is self interference the only mystery of quantum theory. \ There
is in particular the apparent nonlocality of entanglement, and following the
recent rush of \textquotedblleft loop-hole free\textquotedblright\ Bell tests,
it may seem, as has been asserted, that the final nail has been driven into
the coffin of local realism (see, for instance, Wiseman \cite{wiseman}%
).\ \ But no one has as yet come forward to explain these faster-than-light
influences, or to tell us how, if at all, they propagate or how they are to be
reconciled with those fundamental forces of nature for which the speed of
light is a limiting velocity. \ Until more sense can be made of this, these
claims of superluminality are deserving of continuing scrutiny.

Meanwhile, I have shown that in at least one class of physical processes, both
self interference and the Born probabilities can be explained, not merely in a
manner consistent with physical reality, but in accordance with
well-recognized electromagnetic microprocesses that must be suppressed if
measurement is to have the curious nature supposed by SQM. \

\end{document}